\documentclass[10pt, table, xcdraw]{iopart}
\usepackage{multirow}
\usepackage{graphicx}
\usepackage[table,xcdraw]{xcolor}
\usepackage{subcaption}

\begin{document}
\title[Feature matching as improved transfer learning technique for wearable EEG]{Feature matching as improved transfer learning technique for wearable EEG}



\author{Elisabeth R. M. Heremans$^a$\footnote{Corresponding author \\ E-mail address: elisabeth.heremans@kuleuven.be (E. Heremans)}, Huy Phan$^b$, Amir H. Ansari$^a$, Pascal Borz\'ee$^c$, Bertien Buyse$^c$, Dries Testelmans$^c$, Maarten De Vos$^a$}

\address{$^a$ KU Leuven, Department of
Electrical Engineering (ESAT), STADIUS Center for Dynamical Systems, Signal Processing and Data Analytics, Kasteelpark Arenberg 10,
B-3001 Leuven, Belgium}
\address{$^b$ Queen Mary University of London, London E1 4NS,
U.K.}
\address{$^c$ UZ Leuven, Department of Pneumology, Herestraat 49, B-3000 Leuven, Belgium.
}

\begin{abstract}
\textit{Objective:} With the rapid rise of wearable sleep monitoring devices with non-conventional electrode configurations, there is a need for automated algorithms that can perform sleep staging on configurations with small amounts of labeled data. Transfer learning has the ability to adapt neural network weights from a source modality (e.g. standard electrode configuration) to a new target modality (e.g. non-conventional electrode configuration). \textit{Methods:} We propose feature matching, a new transfer learning strategy as an alternative to the commonly used finetuning approach. This method consists of training a model with larger amounts of data from the source modality and few paired samples of source and target modality. 
For those paired samples, the model extracts features of the target modality, matching these to the features from the corresponding samples of the source modality.
\textit{Results:} We compare feature matching to finetuning for three different target domains, with two different neural network architectures, and with varying amounts of training data. Particularly on small cohorts (i.e. 2 - 5 labeled recordings in the non-conventional recording setting), feature matching systematically outperforms finetuning with mean relative differences in accuracy ranging from 0.4\% 
to 4.7\% for the different scenarios and datasets. \textit{Conclusion:} Our findings suggest that feature matching outperforms finetuning as a transfer learning approach, especially in very low data regimes. \textit{Significance:} As such, we conclude that feature matching is a promising new method for wearable sleep staging with novel devices.
\end{abstract}

\maketitle

\ioptwocol

\section{Introduction}
Sleep is crucial to the mental and physical well-being \cite{Siegel2005}, and as such, it comes as no surprise that disturbances in sleep play an important role in a wide variety of diseases \cite{Perez-Pozuelo2020}. Sleep monitoring allows to study sleep and diagnose sleep-wake disturbances. The gold standard for sleep assessment is based on polysomnography (PSG), an overnight recording of multiple physiological signals including electroencephalography (EEG). Such a PSG recording is scored by a trained clinician, who determines the sleep stage corresponding to each 30-second PSG segment, according to developed guidelines \cite{article, Kales1968ASubjects}.\ 

The rapid rise of wearable EEG recording devices has recently started to enable at-home sleep monitoring. These devices will allow to conduct large-scale screenings and longitudinal monitoring to study sleep-wake disturbances and associated diseases on a population level. As such, the emergence of wearables will result in large volumes of sleep data, calling for automated analysis. Moreover, wearable EEG signal modalities are more difficult to interpret for trained clinicians, as the positioning of electrodes differs from the standard EEG electrode placement \cite{Mikkelsen2019a}. Automatic interpretation of these data can alleviate this problem and at the same time reduce the workload of clinicians.\

Currently, reliable methods for automatic sleep staging on wearable data are lacking. Most automated sleep staging methods focus on supervised learning on large annotated PSG datasets, often making use of deep neural network models \cite{ Perez-Pozuelo2020, Phan2018a, Biswal2017, Perslev2021,Phan2020XSleepNet:Staging, Tsinalis2016, Supratak2017, Chambon2018}. However, for sleep staging data from wearable EEG, the only way to get ground truth annotations is through simultaneous acquisition of wearable EEG with full PSG \cite{Mikkelsen2019a}. The manual scoring of the PSG recording can then be used for training automated algorithms on wearable EEG. As this process is costly and time-consuming, the size of annotated wearable EEG datasets is often very small. This greatly limits the performance of supervised learning methods. To compensate for the lack of data, automated sleep staging algorithms for wearable EEG could should exploit information extracted from large, manually labeled datasets with standard EEG modalities. 
After pre-training a model on a large dataset of a standard EEG modality, transfer learning can be used to transfer the learned information to improve the sleep staging performance on a small dataset recorded with a new modality. Previous studies \cite{Phan2019DeepMismatch, Phan2019a, Guillot2021} already used this principle for sleep staging applications, and showed that transfer learning successfully deals with the channel mismatch between different EEG channels (and even across modalities between EEG and EOG). These earlier investigations made use of the simple finetuning approach, in which sleep staging networks were trained on a large labeled dataset of a source modality, and then finetuned on a small dataset of a target modality. In \cite{Phan2020PersonalizedRegularization}, finetuning was used with Kullback-Leibler (KL) divergence regularization for personalization to specific subjects.\ 

Although the finetuning approach is useful, it has some limitations. First, the model forgets the source domain when it is finetuned on the target domain. Therefore, this method does not really allow to study the relationship between the data representations of these different domains. Second, traditional finetuning approaches cannot easily be extended to include unlabeled data, motivating the need to explore alternative methods.\ 

In the field of deep domain adaptation, a specific case of transfer learning \cite{Weiss2016}, domain mapping and domain-invariant feature learning are common strategies to align source and target domain features onto each other \cite{Wilson2018}. Common approaches include domain-adversarial neural networks \cite{Ganin2015} and approaches using the maximum mean discrepancy (MMD) loss \cite{Tzeng2014DeepInvariance, Long2015LearningNetworks}. 
Unsupervised domain adaptation techniques have already demonstrated their potential for EEG-based applications such as emotion recognition and  brain-computer interfaces \cite{Hong2021DynamicClassification, Bao2021Two-LevelRecognition, Zhao2021DeepClassification, Chai2016UnsupervisedRecognition, Li2020DomainSimilarity}. 
These domain adaptation techniques have mainly focused on personalization, cross-session adaptation and domain mismatch between different datasets, with the same or very similar channels recorded in both datasets. In this work, we tackle the challenge of transfer learning between datasets recorded from very different  electrode positions. The difference between the waveforms of the recordings of the source domain and target domain is therefore much larger, which is why we opt for a supervised approach here. 

\begin{figure*}[h]
    \centering
    \includegraphics[width=\textwidth]{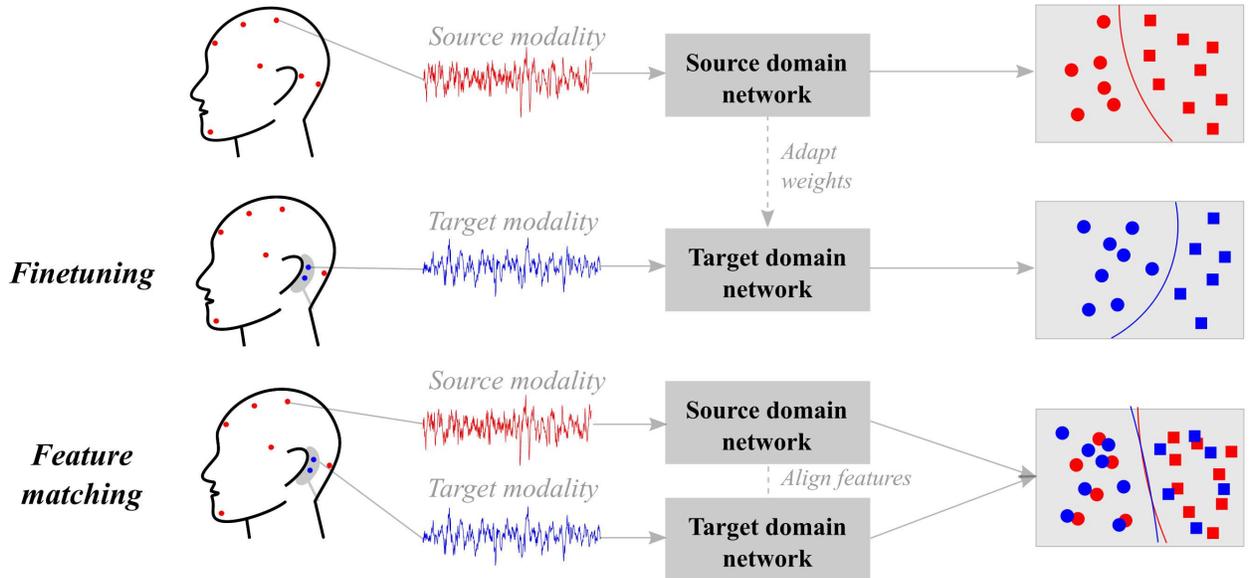}
    \caption{The difference between finetuning and feature matching. Top: (pre-)training of a neural network to perform a classification task on the source domain. Middle: in finetuning, the network weights are adapted by training on the target domain. Bottom: in feature matching, the weights are adapted by training on both the source and the target domain and matching the features of both domains with each other.}
    \label{fig:overview}
\end{figure*}

Combining ideas from supervised  finetuning on the one hand, and unsupervised domain adaptation techniques on the other hand, we propose a novel transfer learning technique (see Fig. \ref{fig:overview}). Our method aligns the feature space of a new EEG modality (the target domain) with that of a standard EEG modality (the source domain). Importantly, we make use of a separate encoding network for the source modality and target modality, a choice which is movitated by the large domain mismatch. Separate encoders allow for more flexibility to learn features separately for both domains. Our method relies on a minimal amount of labels from paired data samples in both modalities. 
Labeled datasets with wearable EEG inevitably have simultaneous recordings of the standard EEG (source modality) and wearable EEG (target modality), because a standard EEG recording is needed to obtain ground truth labels for wearable EEG \cite{Mikkelsen2019a}. Therefore, the use of paired samples from simultaneous recordings is not a limiting factor.

Fig. \ref{fig:overview} illustrates some of the main concepts of this paper that we touched upon in the introduction. The remainder of this paper is structured as follows. Section \ref{materials} describes the datasets used to develop our method and to train and test the classifiers. Section \ref{fm} gives a short introduction on transfer learning, and explains the proposed feature matching method. Section \ref{ssn} discusses the neural network architectures that we use for sleep staging. Then, Section \ref{experiments} reports on the experiments conducted to validate our method and to compare it to the state-of-the-art finetuning approach, and Section \ref{results} shows the obtained results. Finally, Section \ref{discussion} discusses the benefits of transfer learning and the advantages of our method compared to finetuning, and the paper is concluded with Section \ref{conclusion}.

\section{Materials} \label{materials}

\subsection{Source domain} \label{source}
We select the Montreal Archive of Sleep Studies (MASS) database \cite{OReilly2014MontrealResearch} for the source domain, as it is a large, public EEG database. It consists of 200 laboratory-based PSG recordings of 97 men and 103 women aged between 18 and 76 years old, recorded at three different hospital-based sleep laboratories. Sleep stages were scored according to the R\&K guidelines \cite{Kales1968ASubjects} or AASM standard \cite{articleb}. As in \cite{Phan2018a}, we combine the scorings into the five sleep stages of the AASM standard \{W, N1, N2, N3, and REM\} and convert all segments into 30-second ones. 20-second segments are expanded by adding 5 seconds of data before and after the segment. The standard C4-A1 EEG channel from this database is selected as the source domain.

\subsection{Target domain}
For the target domain, we use three different modalities from distinct datasets to test our method extensively in different scenarios.

\subsubsection{MASS – EOG}
The first target domain is the mean electro-oculography (EOG left-right) of the MASS database (see Section \ref{source}) \cite{OReilly2014MontrealResearch}. This target domain allows us to investigate transfer learning from one modality to another one, within one database and population. Forehead electrodes have shown promise for wearable sleep monitoring systems \cite{Younes2016, Lin2017, Rahman2018SleepEOG}. Our approach can thus be used for sleep staging on the EOG signal by itself, which is beneficial for wearable monitoring.

\subsubsection{Surrey – cEEGrid}
The Surrey - cEEGrid database \cite{Sterr2018, Mikkelsen2019a} was recorded at the University of Surrey using the cEEGrid array, a wearable EEG recording device consisting of a flexible printed electrode strip around the ear \cite{Debener2012, Debener2015}. Full-night cEEGrid recordings and PSG recordings were simultaneously collected from 12 healthy adult volunteers. The cEEGrid data were recorded with a wireless SMARTING amplifier (mBrainTrain, Belgrade, Serbia) and a Sony Z1 Android smartphone at a sampling rate of 250 Hz. Manual annotation was based on the PSG \cite{Mikkelsen2019a}. From this dataset, we use the right-ear front-versus-back derivation (FB(R)). This second target domain allows to investigate transfer learning to a dataset of real wearable data, acquired in a completely different sleep laboratory.\

\subsubsection{Leuven – crosshead behind-the-ear} The Leuven - crosshead behind-the-ear sleep database consists of measurements on 28 patients of the sleep laboratory at UZ Leuven. The population is composed of elderly patients with suspicion of sleep apnea. The full PSG was recorded, and an extra EEG electrode was placed behind the right ear, referenced to A1 (located at the left ear). This crosshead behind-the-ear channel simulates a wearable behind-the-ear EEG, which has previously successfully been employed for focal epileptic seizure detection \cite{Becker2021ClassificationDetection, Vandecasteele2020VisualChannels}. In our third target domain, the elderly and diseased population poses an additional challenge, but it is important to validate novel approaches on the target population with suspected sleep problems. Hence, the last target domain  reflects a realistic use case for ambulatory sleep monitoring and allows to investigate transfer learning to a different dataset and modality. This study has the approval of the `Ethics Committee Research UZ/KU Leuven'.

\section{Feature matching} \label{fm}
\subsection{Transfer learning framework}
Transfer learning is the act of improving the performance on a task in a target domain, by using information from a task in a source domain \cite{Weiss2016}. Formally, a domain is defined by an input space $X$ and a marginal probability distribution on that space $P(X)$: $D=\{X,P(X)\}$. A task is defined as its label space $Y$ and predictive function $f(.)$ that projects $X$ onto $Y$: $T=\{Y, f(.)\}$. In classical machine learning, the domains and tasks of training set and test set are assumed to be the same. When a mismatch between either the tasks or the domains occurs, transfer learning aims to account for this discrepancy. Formally, when $T_S \neq T_T$ or $D_S \neq D_T$ (where subscript $S$ and $T$ indicate source and target, respectively), transfer learning improves the predictive function $f_T(.)$ using information from $T_S$ and $D_S$ \cite{Weiss2016, Zhuang2021ALearning}.\

Applying these concepts to sleep stage classification on wearable EEG recordings, the source domain consists of a standard EEG channel of a large public sleep database. We aim to transfer knowledge from this domain to the target domain consisting of a small database with a non-traditional EEG channel. The task in both domains is sleep stage classification into the five sleep stages.

\subsection{Feature matching method}
Similar to the finetuning approach, we first pre-train a sleep staging network on a large sleep database, i.e. the source domain. 
Then, we use our novel feature matching method as a transfer learning approach to adapt this network to the small database with a new modality, i.e. the target domain. Whereas the finetuning approach only uses the labels of the target modality, our feature matching approach 
also exploits the correspondence between the simultaneously recorded data of both the source modality and target modality. We explicitly match the extracted feature vectors of the target modality with those of the source modality for the corresponding samples.

The reasoning behind this approach, is that through pre-training on a large source dataset, we learn features from the source domain which are superior to those we can learn from a very small target dataset. By minimizing the distance between the features of the source modality and target modality, we use information from the source domain to improve the target domain features. Feature vectors represent a precise location in the feature space, whereas a label only designates a general area in the feature space.

\begin{figure*}[h]
    \centering
    \includegraphics[width=0.8\textwidth]{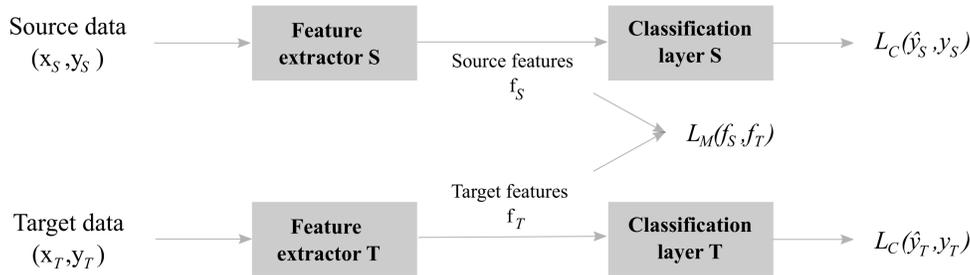}
    \caption{Feature matching consists of concurrently training two networks, while matching the feature representations.}
    \label{fig:featurematching}
\end{figure*}

Each sleep staging network is conceptually split into two components: a feature extractor consisting of all layers but the last one, and the last layer itself which is the classification layer. Both of these are trained end-to-end as one network, but perform different tasks. The feature matching method, as illustrated in Fig. \ref{fig:featurematching}, consists of two steps:

\begin{enumerate}
\item Initialization: make two duplicates of a sleep staging network architecture (see Fig. \ref{fig:featurematching}), and initialize them both with the network weights pre-trained on the source domain,
\item Adaptation of the two parallel networks, with both networks trained on different modalities. The first network, the source network (consisting of feature extractor S and classification layer S in Fig. \ref{fig:featurematching}), is further trained on the source modality. The second network, the target network (feature extractor T and classification layer T), is finetuned on the target modality.
\end{enumerate}

Both the source and target modality networks get separately trained to classify sleep stages (network predictions $\hat{y}$) with data ($x$) and labels ($y$) from their respective modality. At the same time, their feature extractors are trained to minimize the MSE between extracted features ($f$) of corresponding samples of the source modality and target modality. This training process is what we refer to as feature matching. The combination of both the source modality network and target modality network thus gets trained with the following loss function:
\begin{eqnarray}
\fl 
L = L_C(\hat{y}_S, y_S) + L_C(\hat{y}_T, y_T) \nonumber \\ 
+ \lambda_1 L_M(f_S ,f_T) + \lambda_2 L_2
\label{loss}
\end{eqnarray}
in which the subscript $S$ and $T$ designate the source modality and target modality, respectively. $L_C$ is the cross-entropy or classification loss, $L_M$ is the matching loss between the features of paired samples of the source modality and the target modality, computed with the MSE, and $L_2$ is the $L_2$-norm of the network weights. The hyperparameter $\lambda_1$ determines the weight of the matching loss relative to the classification losses, and $\lambda_2$ determines the weight of the $L_2$ regularization term.

Alternatively to using the MSE loss between paired samples, the MMD loss could also be used to match the source and target features \cite{Tzeng2014DeepInvariance, Long2015LearningNetworks}. However, this measure acts on a distribution level instead of using paired samples, not taking full advantage of the available knowledge in the case of simultaneous EEG recordings. In our experiments, the accuracy obtained with the MSE loss systematically outperformed the accuracy obtained when using the MMD loss.

\section{Sleep staging networks} \label{ssn}
The experiments are performed with two different neural network architectures for sleep staging: a compact 3-layer attention-based recurrent neural network (ARNN) \cite{Phan2018a} on the one hand, and a state-of-the-art sleep staging network, SeqSleepNet \cite{Phan2018a} on the other hand. Both are illustrated in Fig. \ref{fig:networks}. The networks are trained to classify 30-second segments of recorded data into the five sleep stages according to the AASM standard \cite{articleb}: Wake, sleep stages N1, N2 and N3, and REM sleep. They can both cope with different numbers of input channels, but in this work, we focus on single-channel sleep staging. All the data are pre-processed in the same way before being presented to these networks.
\subsection{Pre-processing}
All the recorded signals are filtered between 0.3 and 40 Hz and resampled to 100 Hz. Then, every recording gets transformed to its logarithmically scaled time-frequency spectrum, and this spectrogram is normalized to zero mean and unit standard deviation. 
\subsection{Attention-based recurrent neural network}
The simple ARNN (Fig. \ref{fig:networks}(a)) \cite{Phan2018a} follows the classical one-to-one classification scheme, meaning it takes a single segment as input and outputs its corresponding sleep stage. It consists of three layers. The first layer is a filterbank layer that filters the frequency dimension with learned weights. Then follows a bidirectional recurrent neural network (biRNN) implemented with a gated recurrent unit (GRU) cell. This layer allows for sequential modelling of the temporal information within a 30-second segment. The third and last layer is an attention layer, which combines the vectors extracted by the biRNN into one vector, the final feature representation of the segment. Classification is performed by passing this feature vector through a fully connected layer with a softmax activation. The network is trained end-to-end by minimizing the cross-entropy loss.
\subsection{SeqSleepNet}
\begin{figure*}
     \centering
     \begin{subfigure}[b]{0.267\textwidth}
         \centering
         \includegraphics[width=\textwidth]{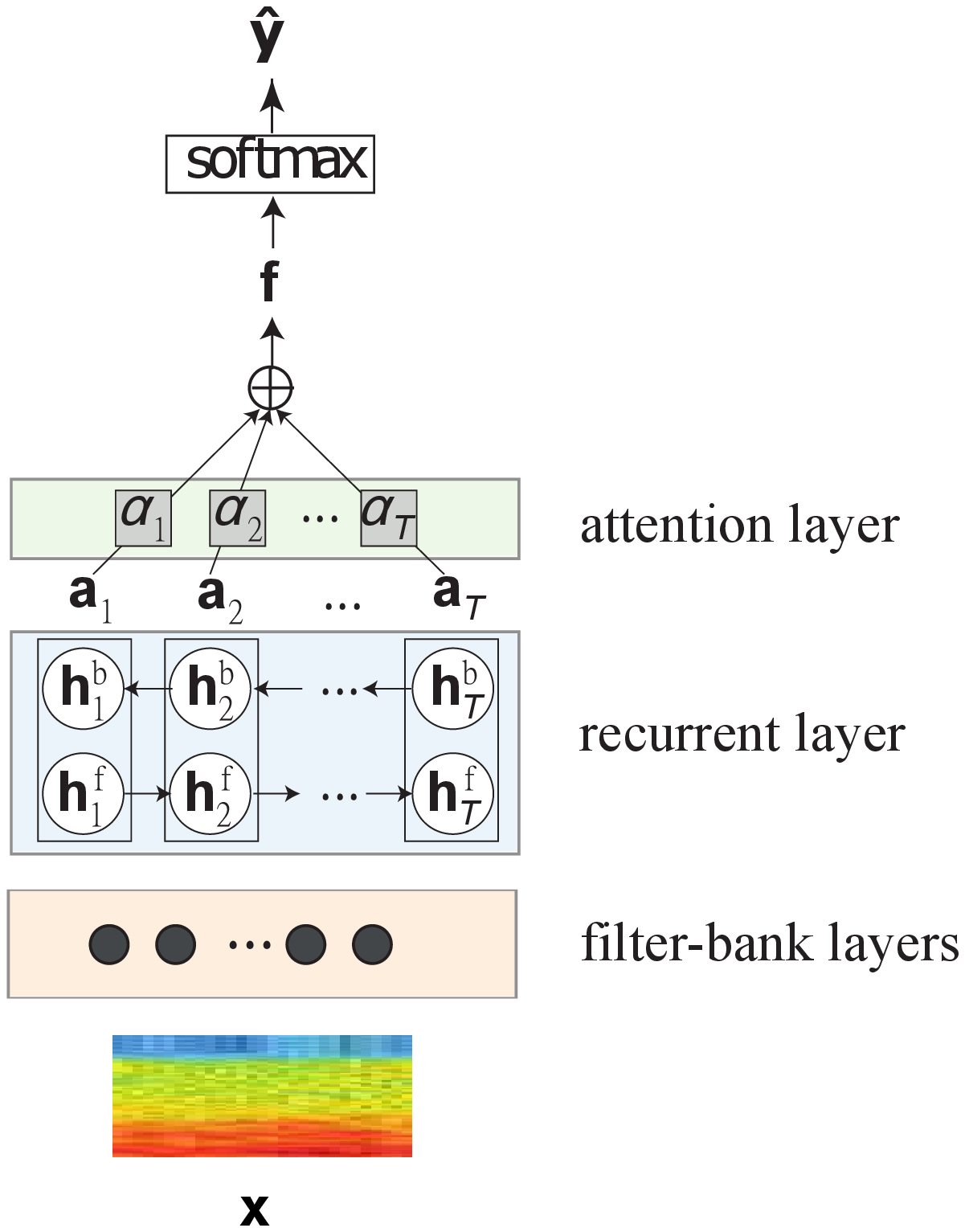}
         \caption{}
         \label{fig:arnn}
     \end{subfigure}
     \begin{subfigure}[b]{0.70\textwidth}
         \centering
         \includegraphics[width=\textwidth]{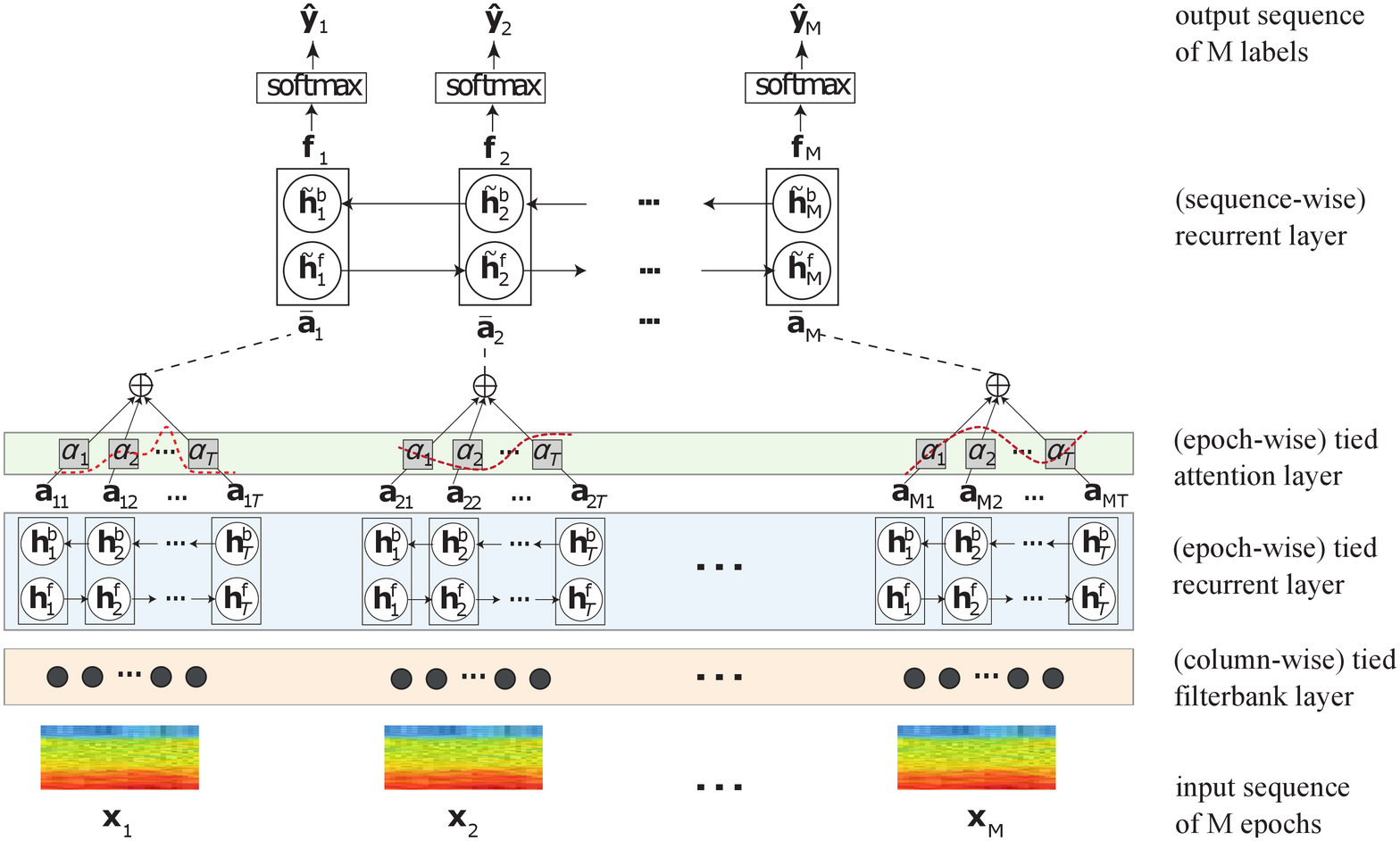}
         \caption{}
         \label{fig:seqsleepnet}
     \end{subfigure}
        \caption{Illustration of the two sleep staging networks used in this study. (a) The Attention-based Recurrent Neural Network, performing single segment-based sleep staging, (b) SeqSleepNet, designed for sequence-to-sequence classification. Both figures are adapted from \cite{Phan2018a}.}
        \label{fig:networks}
\end{figure*}

SeqSleepNet (Fig. \ref{fig:networks}(b)) \cite{Phan2018a} follows a many-to-many classification scheme, so it takes multiple segments as input and predicts all of the corresponding sleep stages at once. It transforms a sequence of $M$ segments into the corresponding sequence of $M$ sleep stages. In this study, we use a sequence length $M=10$. The ARNN architecture is used as the first block of SeqSleepNet, outputting a single feature vector per segment. Then, the feature vectors of a whole sequence of segments are presented to a second biRNN layer acting at a sequence level, which models the temporal relationship between the segments. This sequence-level biRNN layer is implemented in the same way as the segment-level biRNN layer. It takes a sequence of $M$ input feature vectors and transforms it to a sequence of $M$ output feature vectors. Those $M$ output vectors are then classified into $M$ sleep stages by a fully connected layer with a softmax activation. The network is trained in an and-to-end manner, by minimizing the cross-entropy loss averaged over the $M$ segments.\ 

To train SeqSleepNet on all the possible sequences in a dataset, we sample sequences from the dataset with a shift of one segment, i.e. an overlap of $M-1$ segments. At test time, sampling the test set with the same shift results in an ensemble of $M$ predictions for every segment. These predictions are aggregated by summing the logarithmic posterior probabilities over the ensemble. 
For further details of both networks, see \cite{Phan2018a}.


\subsection{Parameters and settings}
The ARNN and SeqSleepNet are both implemented with the \textit{Tensorflow} framework \cite{10.5555/3026877.3026899}.
The networks are parametrized the same way as in the original paper \cite{Phan2018a}, and are trained with the Adam optimizer and a learning rate of $1e-4$.

\section{Experiments} \label{experiments}
\subsection{Transfer learning scenarios}

In order to validate our feature matching method and compare it to the state-of-the-art finetuning approach, we examined different transfer learning scenarios. In every scenario, the model was pre-trained on the C4-A1 derivation of a large dataset, and then adapted to overcome the channel mismatch with different target modalities from small datasets, each containing both the C4-A1 source derivation and the target modality. \

We applied transfer learning to three different target modalities: EOG, cEEGrid, and crosshead behind-the-ear. In every scenario, transfer learning was performed with the classical finetuning approach, and with the novel feature matching method. In addition to the basic finetuning approach, finetuning with KL-divergence regularization as introduced in \cite{Phan2020PersonalizedRegularization} was also added as an extra baseline for comparison. Each of the experiments was performed once using the compact ARNN network, and once with SeqSleepNet. We also compared the transfer learning performances with simple training from scratch on the target domain, and with directly evaluating the networks trained on the source domain (we call this `direct transfer').

\subsection{Experimental setup} \label{expsetup}

For every learning scenario, we trained the models on different amounts of target modality data to investigate how both transfer learning methods perform on smaller and larger datasets. Every experiment was performed as a cross-validation on the target dataset to obtain average performance measures. Smaller subsets within each fold’s training set were made, to construct training sets of different sizes for transfer learning. The sizes of these training sets for transfer learning were of 10, 5 and 2 recordings. Table \ref{datasetdescription} shows the amount of recordings per dataset, the subdivision into a training set, test set and validation set for every fold, and the number of recordings in the subsets. Note that depending on the size of the dataset, there are multiple different possible subsets within one fold's training set. For the subset of 10 recordings, we picked 10 recordings within each fold. Within those 10 recordings, we made 2 non-overlapping subsets of 5 recordings and 5 non-overlapping subsets of 2 recordings. Hence, for every fold, we use 1 training dataset of 10 recordings, 2 training sets of 5 recordings and 5 training sets of 2 recordings. The performance values are averaged over all the folds and subsets within the folds. Since the variability in performance is higher when smaller training sets are used, it makes sense to split the data into more different training sets when the training sets are smaller, to get a representative performance value.

\begin{table*}[h]
\caption{\label{datasetdescription}The datasets and their subdivision into training, validation and test sets for every cross-validation fold. `Training set' designates the total training set for a fold, and `training subsets' designates the training sets used for transfer learning.}
\lineup
\begin{tabular}{@{}llllllll}
\br
Dataset          & Channel        & \multicolumn{5}{c}{Number of recordings}                                           & Nb. \\  \ns &&\crule{5} & folds \\
                 &                & Total & Training & Training & Val. & Test &          \\ \ns
               &&& set& subsets & set & set &  \\ \mr
MASS-C4          & C4-A1          & 200   & 180                &                                 & 10       & 10       & 20       \\
MASS-EOG         & EOG            & 200   & 180                & 10/5/2                          & 10       & 10       & 20       \\
Surrey-cEEGrid   & cEEGrid        & \012    & \010                 & 10/5/2                          & \01        & \01        & 12       \\
Leuven-crosshead & Right ear-A1 & \028    & \024                 & 10/5/2                          & \02        & \02        & 14  \\  \br
\end{tabular}
\end{table*}

The first set of experiments was carried out with the EOG from the MASS dataset as the target domain. In this case, the dataset of the source domain and target domain are actually the same. Pre-training was performed with 20-fold cross-validation on the C4-A1 recordings of the MASS dataset, with 180 recordings as a training set in every fold. Then, transfer learning was performed with the same 20-fold cross-validation scheme, this time only using 10 EOG and C4-A1 recordings in each iteration of the cross-validation to simulate smaller datasets. Those 10 recordings were further subdivided into the aforementioned subsets for transfer learning (one subset of 10 recordings, two subsets of 5 recordings and five subsets of 2 recordings).

The experiments on the Surrey-cEEGrid dataset and the Leuven-crosshead behind-the-ear dataset follow a simpler scheme. In both of these cases, we pre-trained the networks on all 200 C4-A1 recordings of the MASS dataset, except 10 recordings used as a validation set. Then, we performed a cross-validation on the two respective target datasets, with a further subdivision into the training subsets. For further details of the cross-validation and subsets, we refer to Table \ref{datasetdescription}.

\subsection{Training parameters}
\begin{figure}[h]
    \centering
    \includegraphics[width=0.45\textwidth]{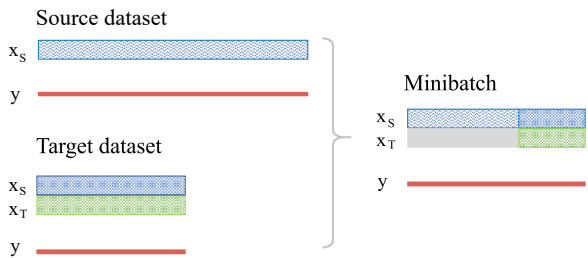}
    \caption{Illustration of a minibatch for training the feature matching approach. $x_S$ indicates data of the source modality, and $x_T$ indicates data of the target modality. $y$ indicates the labels.}
    \label{fig:minibatch}
\end{figure}

For feature matching, the minibatches are constructed in the following manner, illustrated in Fig. \ref{fig:minibatch}. For the source modality network to retain its sleep staging capabilities on the source modality, it is presented with labeled source modality data from both the source dataset (the MASS dataset in this case) and the target dataset during the feature matching process. In the meantime, the target modality network is presented with only target modality samples of the target dataset. The matching loss can only be computed for the paired samples that have both modalities, i.e. the samples of the target dataset. Every minibatch consists partly of data from the source modality (used for $L_C(\hat{y}_S, y_S)$ in the loss function \ref{loss}), and partly of target modality samples with their corresponding source modality samples (used for $L_M(f_S ,f_T)$ and $L_C(\hat{y}_T, y_T)$ in \ref{loss}). In comparison, it should be noted that for simple finetuning, the minibatches are constructed using only target modality samples of the target dataset, as in this case, the model only trains with the classification loss on the target modality samples.

Networks are always pre-trained for 10 epochs and transfer learning (feature matching or finetuning) is performed for 20 epochs. During training, networks are evaluated on the validation set after every 200 iterations, and the best-performing network on the validation set is retained for evaluation on the test set. This acts as a regularization approach like early stopping. For pre-training and finetuning, the minibatches are of size 32. For feature matching, the size of minibatches ($N_{mb}$) is not constant across all training scenarios. The minibatches contain both 8 samples from the target dataset ($N_{td,mb}=8$) and a variable, larger number of samples from the source dataset ($N_{sd,mb}$). The number of samples from the source dataset in a minibatch depends on the proportion in size of the two datasets ($N_{sd}/N_{td}$), so that all the training samples of both datasets pass through the network once in every epoch: $N_{mb} =N_{td,mb}+ N_{sd,mb} = 8+ 8* N_{sd}/N_{td}$. In every training step, the classification losses and matching loss are computed by summing over the minibatch. As the matching loss $L_M$ is summed over only 8 samples of the minibatch, the matching loss weight $\lambda_1$ is fixed to $N_{mb}/8$ to give it the same relative importance as the source classification loss. 

\section{Results} \label{results}
\subsection{Baseline performance on the source domain} 
First, to put our results in perspective, we show the network performances on the source domain. Training on the best possible EEG channel of a large dataset acquired from healthy subjects with standard hospital equipment should give an upper limit to what the networks can achieve with single-channel data. 20-fold cross-validation is performed on the source domain, with 180 recordings in the training set, 10 recordings as validation set and 10 recordings as test set for every fold. Table \ref{res_eog} shows the sleep staging performance of SeqSleepNet and ARNN on this dataset. It reports the accuracy, Cohen’s kappa ($\kappa$) and weighted F1-score (wF1), as mean ± standard error over the 20 folds. The accuracy obtained with SeqSleepNet on single-channel data is 83.9\%, which is in line with the results for multiple channels in \cite{Phan2018a}.



\subsection{Finetuning and feature matching performance on the target domains}
\begin{figure*}[h]
\centering
    \includegraphics[width=0.95\textwidth]{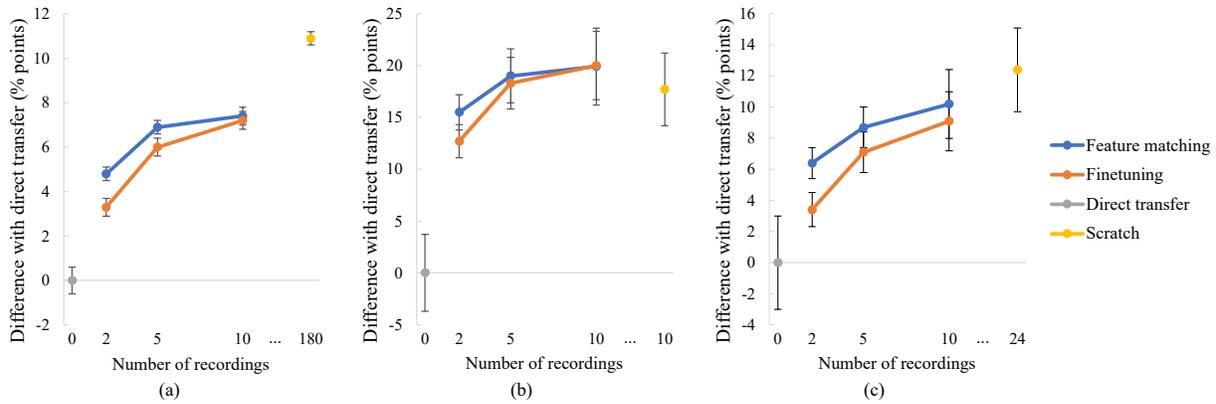}
    \caption{Visual representation of some important results from Table \ref{res_eog}. Absolute difference in accuracy of feature matching, finetuning and training from scratch with respect to direct transfer. Results are shown using SeqSleepNet as a network, and for the three target datasets: (a) MASS - EOG, (b) Surrey - cEEGrid, (c) Leuven - crosshead behind-the-ear.}
    \label{fig:results}
\end{figure*}

We assess the performance of both transfer learning methods and compare it with direct transfer and training from scratch on the target domain. Table \ref{res_eog} shows the results for the three target domains: the EOG of the MASS database, the cEEGrid channel of the Surrey dataset, and the crosshead behind-the-ear channel of the Leuven dataset. The mean ± standard error is computed over all the cross-validation folds and subsets within the folds as defined in Section \ref{expsetup}. Fig. \ref{fig:results} highlights some of the most important results, visualizing the difference in performance of the different methods for the three target domains.\

The baseline performance on the large source domain is clearly higher compared to the performances obtained from training on smaller sized datasets of the target domains in Table \ref{res_eog}. For every target domain, direct transfer performs worse than any other method tested.
Training from scratch on the complete dataset performs worse than transfer learning on subsets when the dataset is small (e.g. the Surrey - cEEGrid dataset), but it performs better when the dataset is large (e.g. the MASS - EOG dataset). 
When comparing the two transfer learning techniques, feature matching always outperforms finetuning when 2 recordings of the target modality are used. When 5 recordings are used, feature matching also outperforms finetuning in most cases. When using 10 recordings, the difference is very small, so we could state that both approaches obtain a similar performance.

\begin{table*}[h]
\small
\centering
\caption{\label{res_eog}Sleep staging performance of ARNN and SeqSleepNet on source and target domains, using feature matching, finetuning, finetuning with KL-divergence regularization, direct transfer and training from scratch. Mean ± standard error is set out for the accuracy (acc.), Cohen’s kappa ($\kappa$) and weighted F1-score (wF1). \# is the number of recordings in the training set for every scenario. The grey highlight indicates whether feature matching or finetuning obtains the best performance for each scenario.}
\lineup
\begin{tabular}{@{}llcccccc}
\br
\centre{8}{Source domain: MASS - C4-A1 dataset}\\ \mr
\# &     Method     &\multicolumn{3}{c}{ARNN}          & \multicolumn{3}{c}{SeqSleepNet}   \\ \ns &&\crule{3}&\crule{3}\\

&&Acc      & $\kappa$           & wF1      & Acc      & $\kappa$          & wF1      \\ \mr
180&Scratch&80.1±0.5 & 0.717±0.007 & 79.6±0.4 & 83.9±0.4 & 0.769±0.006 & 83.6±0.4 \\ \ms
\br
\centre{8}{Target domain 1: MASS - EOG dataset} \\ \mr
\# &     Method             & \centre{3}{ARNN}           & \centre{3}{SeqSleepNet}   \\\ns && \crule{3}&\crule{3} \\
                      &                  & Acc       & $\kappa$           & wF1      & Acc      & $\kappa$           & wF1      \\ \mr
\multirow{3}{*}{2}  & Feature matching   & \textbf{73.3±0.3} & \textbf{0.617±0.004} & \textbf{72.8±0.3} & \textbf{77.6±0.3}    & \textbf{0.680±0.004} & \textbf{77.1±0.3} \\

                    & Finetuning         & 71.9±0.4          & 0.603±0.005          & 71.7±0.4          & 76.1±0.4             & 0.656±0.005          & 75.5±0.4          \\
                     & Finetuning with KL & 73.0±0.3          & 0.614±0.004          & 72.5±0.3          & 77.5±0.3             & 0.671±0.004          & 76.4±0.3          \\                   \ms
\multirow{3}{*}{5}  & Feature matching   & \textbf{74.9±0.4} & \textbf{0.642±0.005} & \textbf{74.5±0.4} & \textbf{79.7±0.3}    & \textbf{0.709±0.005} & \textbf{79.2±0.4} \\

                    & Finetuning         & 74.6±0.4          & 0.640±0.005          & \textbf{74.5±0.4} & 78.8±0.4             & 0.697±0.006          & 78.6±0.4          \\
                     & Finetuning with KL & 74.7±0.4          & 0.639±0.005          & 74.2±0.3          & 79.2±0.3             & 0.697±0.005          & 78.4±0.4          \\                   \ms
\multirow{3}{*}{10} & Feature matching   & \textbf{76.1±0.4} & \textbf{0.659±0.006} & 75.7±0.4          & \textbf{80.2±0.4}    & \textbf{0.715±0.006} & 79.7±0.4          \\

                    & Finetuning         & 76.0±0.4          & \textbf{0.659±0.006} & \textbf{75.8±0.4} & 80.0±0.4             & 0.714±0.005          & \textbf{79.8±0.3} \\
                     & Finetuning with KL & 75.7±0.4          & 0.652±0.006          & 75.2±0.4          & 80.0±0.4             & 0.709±0.006          & 79.2±0.5          \\                   \ms
180                 & Scratch            & 79.4±0.3          & 0.706±0.004          & 78.4±0.3          & 83.7±0.3             & 0.766±0.004          & 83.3±0.3          \\
0                   & Direct transfer    & 70.1±0.6          & 0.561±0.008          & 69.4±0.7          & 72.8±0.6             & 0.592±0.009          & 70.6±0.7         
\\
\ms \br 
\centre{8}{Target domain 2: Surrey - cEEGrid dataset} \\ \mr

\# &          Method        & \multicolumn{3}{c}{ARNN}          & \multicolumn{3}{c}{SeqSleepNet}   \\\ns && \crule{3}&\crule{3}\\
                      &                  & Acc      & $\kappa$           & wF1      & Acc      & $\kappa$           & wF1      \\ \mr
\multirow{3}{*}{2}          & Feature matching   & \textbf{63.9±1.5} & \textbf{0.478±0.021} & \textbf{60.2±1.8} & \textbf{66.9±1.7}    & \textbf{0.526±0.023} & 62.3±2.0          \\

                            & Finetuning         & 61.5±1.4          & 0.444±0.021          & 58.2±1.7          & 64.1±1.6             & 0.513±0.019          & 61.6±1.8          \\
                            & Finetuning with KL & 62.6±1.4          & 0.460±0.021          & 59.5±1.8          & 66.0±1.2             & 0.514±0.017          & \textbf{63.7±1.2} \\                    \ms
\multirow{3}{*}{5}          & Feature matching   & \textbf{68.4±2.5} & \textbf{0.543±0.036} & \textbf{64.8±2.9} & \textbf{70.4±2.6}    & 0.577±0.034          & 67.3±2.9          \\
                            & Finetuning         & 66.7±2.2          & 0.521±0.032          & 63.7±2.4          & 69.7±2.5             & \textbf{0.584±0.033} & 67.5±2.9          \\                            & Finetuning with KL & 66.6±2.2          & 0.517±0.033          & 63.6±2.7          & 69.1±2.5             & 0.572±0.032          & \textbf{67.8±2.4} \\
                    \ms
\multirow{3}{*}{10}         & Feature matching   & \textbf{69.1±3.1} & \textbf{0.556±0.043} & 65.9±3.5          & 71.3±3.7             & \textbf{0.605±0.040} & 68.6±3.6          \\
                            & Finetuning         & 68.4±2.7          & 0.548±0.039          & \textbf{66.1±3.0} & \textbf{71.4±3.3}    & 0.597±0.046          & \textbf{70.5±3.2} \\     & Finetuning with KL & 68.1±3.1          & 0.542±0.045          & 65.7±3.8          & 70.6±3.2             & 0.577±0.047          & 68.9±3.3          \\
                    \ms
10                          & Scratch            & 66.7±3.1          & 0.524±0.043          & 63.8±3.3          & 69.1±3.5             & 0.575±0.041          & 66.3±3.6          \\
0                           & Direct transfer    & 58.2±2.3          & 0.410±0.031          & 56.7±2.4          & 51.4±3.7             & 0.375±0.032          & 50.4±3.1         
\\ \ms \br
\centre{8}{Target domain 3: Leuven - crosshead behind-the-ear dataset} \\ \mr
\#                  &      Method            & \multicolumn{3}{c}{ARNN}          & \multicolumn{3}{c}{SeqSleepNet}   \\\ns && \crule{3}&\crule{3}\\
                    &                  & Acc      & $\kappa$           & wF1      & Acc      & $\kappa$           & wF1      \\\mr
\multirow{3}{*}{2}      & Feature matching   & \textbf{63.1±1.0} & 0.484±0.014          & 62.4±1.1          & \textbf{67.5±1.0}    & \textbf{0.544±0.014} & 66.3±1.1          \\
                            & Finetuning         & 61.6±1.1          & 0.466±0.014          & 60.7±1.1          & 64.5±1.1             & 0.508±0.015          & 63.6±1.2          \\
                            & Finetuning with KL & \textbf{63.1±1.0} & \textbf{0.489±0.014} & \textbf{63.0±1.1} & 66.9±1.1             & 0.541±0.014          & \textbf{66.5±1.2} \\\ms
\multirow{3}{*}{5}      & Feature matching   & \textbf{66.1±1.4} & \textbf{0.522±0.020} & \textbf{65.3±1.4} & \textbf{69.8±1.3}    & \textbf{0.573±0.018} & \textbf{68.8±1.4} \\
                            & Finetuning         & 64.6±1.4          & 0.504±0.020          & 64.0±1.4          & 68.2±1.3             & 0.552±0.019          & 67.8±1.4          \\                            & Finetuning with KL & 65.0±1.5          & 0.512±0.020          & 64.9±1.5          & 68.1±1.5             & 0.554±0.021          & 68.2±1.6          \\\ms
\multirow{3}{*}{10}     & Feature matching   & 67.0±2.2          & 0.531±0.032          & 65.9±2.3          & \textbf{71.3±2.2}    & \textbf{0.592±0.031} & \textbf{70.4±2.2} \\
                            & Finetuning         & \textbf{67.1±2.1} & \textbf{0.534±0.030} & 66.1±2.2          & 70.2±1.9             & 0.576±0.028          & 69.9±2.0          \\
                            & Finetuning with KL & 66.9±2.0          & \textbf{0.534±0.028} & \textbf{66.6±2.2} & 69.7±2.3             & 0.575±3.3            & 69.7±2.5          \\             \ms
24                          & Scratch            & 67.5±2.4          & 0.541±0.033          & 65.5±2.5          & 73.5±2.7             & 0.618±0.039          & 70.9±2.9          \\
0                           & Direct transfer    & 60.2±2.3          & 0.452±0.031          & 60.9±2.5          & 61.1±3.0             & 0.488±0.036          & 61.3±3.3         \\
\br
\end{tabular}
\end{table*}

\subsection{Visualizing the feature spaces}
\begin{figure*}[h]
    \centering
    \includegraphics[width=.8\textwidth]{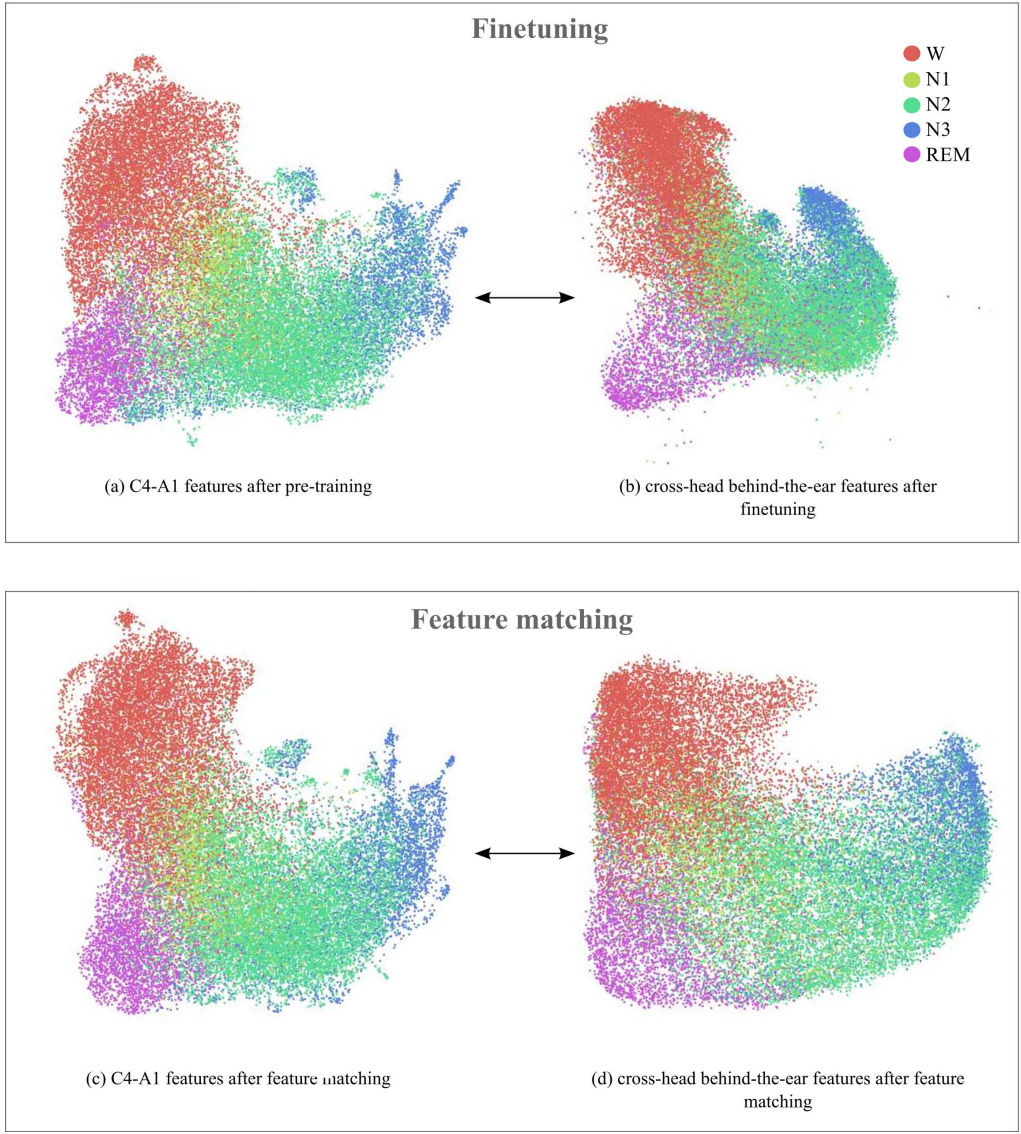}
    \caption{UMAP visualization of the C4-A1 and crosshead behind-the-ear modality features (both of the Leuven dataset), learned by SeqSleepNet, before and after transfer learning on 5 recordings of the Leuven dataset. Each point in a cloud represents the feature vector of one 30-second segment, and the different colors are different sleep stages. (a) C4-A1 features after pre-training on the source domain, (b) crosshead behind-the-ear features after finetuning on the target domain, (c) C4-A1 features after feature matching, (d) crosshead behind-the-ear features after feature matching.}
    \label{fig:umap}
\end{figure*}

To better understand the difference between finetuning and feature matching, we can visualize the features learned by the sleep staging network using both technique. For this purpose, we use the Uniform Manifold Approximation and Projection (UMAP) technique \cite{McInnes2018UMAP:Projection}. This projects the features from their high-dimensional space to two dimensions, and thus allows every sample to be plotted as a point in a 2D plane. Fig. \ref{fig:umap} shows the feature spaces learned by SeqSleepNet when performing transfer learning to the crosshead behind-the-ear recordings of the Leuven dataset. It plots the UMAP projections of both C4-A1 and the crosshead behind-the-ear modality. Fig. \ref{fig:umap}(a) shows the feature space of the source modality (C4-A1) after pre-training, and Fig. \ref{fig:umap}(b) shows how the feature space changes shape after finetuning on the target modality (crosshead behind-the-ear). Then, Fig. \ref{fig:umap}(c)-(d) show the feature spaces of both modalities after feature matching. As feature matching acts on both the source modality and the target modality, we plot both the learned C4-A1 and crosshead behind-the-ear features in this case.

\section{Discussion} \label{discussion}

In the present study, we propose feature matching, a novel transfer learning approach designed to transfer knowledge from a standard EEG set-up to a database of a new EEG recording modality with potentially large differences in waveforms. Using a sleep staging task, we compare this method to finetuning, the state-of-the-art transfer learning approach, and to the baseline approaches of direct transfer and training from scratch. In Table \ref{res_eog}, we validate our method for diverse scenarios: we use two neural network architectures, and three different target modalities of distinct datasets acquired at different locations and sleep laboratories, and recorded from different populations. In addition, we analyze the effect of the transfer learning methods with varying sizes of target domain training datasets. 

The need for adapting the models to the target domains is clear from the low accuracies obtained with direct transfer. All the transfer learning scenarios, even using as little as two recordings of the target modality, achieve higher accuracies than direct transfer. Certain scenarios require more adaptation than others. In cases where the performance with direct transfer is the lowest, the relative (and absolute) gains obtained from transfer learning are clearly larger. The relative percent difference in accuracy between feature matching on two recordings and direct transfer is only 6.6\% for the MASS - EOG target domain, but 10.5\% on the Leuven - crosshead behind-the-ear domain and 30.2\%  on the Surrey - cEEGrid domain, using the SeqSleepNet architecture. The EOG, from the same dataset as the source domain, and therefore recorded with the same technology and from the same population, clearly requires the least adaptation. The other two target domains require more adaptation as they are recorded with different (wearable) devices, from different populations. This proves that adaptation techniques are necessary to deal with channel mismatch and other types of mismatch, and both the transfer learning strategies fulfill that goal.

For all the target domains, the models are also trained from scratch on the complete dataset. The performances achieved with this approach strongly depend on the size of the dataset. For the EOG target domain, the network trained from scratch performs better than all other methods, because it uses 180 recordings (like the baseline network trained on the source domain). For the crosshead behind-the-ear target domain, transfer learning on 10 recordings achieves a comparable performance to training from scratch on 24 recordings. For the cEEGrid target domain, training from scratch on 10 recordings performs worse than transfer learning on 10 recordings, and comparably to transfer learning with 5 recordings. The amount of training data in the cEEGrid dataset, 10 recordings, is thus clearly too small to train the large amount of model parameters and achieve good generalization without relying on transfer learning. These results again show the usefulness of transfer learning in small data regimes, as it requires less data to achieve similar performances to training from scratch.

When we compare the two transfer learning techniques, feature matching and finetuning, the main difference between the methods lies in the minimization of the distance between source features and target features of the same samples. 
Fig. \ref{fig:umap} demonstrates the effect this has on the learned features for both modalities. 
Feature matching aligns the feature spaces of the two modalities (Fig. \ref{fig:umap}(c) and (d)), whereas finetuning adapts the feature space of the target modality (Fig. \ref{fig:umap}(b)) without aiming to match the source modality. 

In terms of classification performance, feature matching clearly has an advantage over finetuning in the smaller data regimes (the training scenarios on 2 and 5 recordings). When 10 (or more) recordings of the target domain are used, the two methods perform on par with each other. The less data are available from the target domain, the more the additional information from the source domain helps the model to achieve a better performance. We can understand this advantage as follows. First, a position in the feature space contains more precise information than a sleep label. The label only tells the network which region in the feature space the feature vector belongs to, whereas the use of the source feature vector adds the precise location in that feature space. Second, the feature matching technique allows to remember what was learned from the source domain, and matches the features of the target domain onto features of the source domain. As the source network is trained on a much larger dataset, the source features are of superior quality, with a better separation of the sleep stages than the target features. Aligning the target features onto those superior source features thus aids the target network to achieve a better performance as it acts as a form of regularization and implicitly exploits the information extracted through training on a much larger dataset. 

With the addition of the extra baseline method of finetuning with KL-divergence regularization, we investigate whether part of the performance gap between finetuning and feature matching can be bridged with the addition of strong regularization in the finetuning approach. Indeed, the finetuning approach with KL-divergence regularization generally performs better than the basic finetuning approach in the training scenario of 2 recordings, but it still mostly achieves lower performances than feature matching. This result supports the notion that feature matching adds more value than just regularization.


Our implementation of feature matching makes use of the MSE loss between paired samples, exploiting the availability of simultaneous recordings of the source and target modality. 
In other application areas where simultaneous recordings of a source domain and target domain might not be available, we can adapt the technique to align both domains without using the explicit correspondence between samples. The general feature matching idea and structure (Fig. \ref{fig:featurematching}) does not change, but the matching loss can be implemented differently, for example with the MMD loss.

A minor disadvantage of the feature matching technique compared to finetuning is the longer training time. As the feature matching structure consists of two networks instead of one, and requires training both of those structures with data from two modalities and datasets instead of one, it requires more computational power and has a longer training time (about 8 to 9 times longer in our set of experiments) than the transfer learning technique. 

\section{Conclusion} \label{conclusion}
This work presents feature matching, a novel transfer learning technique for deep neural networks performing sleep staging tasks. Our method is specifically tailored towards adapting sleep staging networks from standard EEG channels to new, non-standard EEG channels.  Contrary to existing domain adaptation methods for EEG, this method explicitly uses the correspondence between simultaneous recordings of a standard and a non-standard EEG channel to improve the sleep staging performance on small wearable datasets. 
As such, in small data regimes, feature matching significantly outperforms finetuning, the standard transfer learning technique for this application. We conclude that this feature matching method has a lot of promise to improve sleep staging performances in small datasets with non-standard EEG modalities. 

\section*{CRediT authorship contribution statement}
\textbf{Elisabeth R. M. Heremans:} Conceptualization, Methodology, Software, Validation, Formal Analysis, Resources, Writing - Original Draft, Writing - Review \& Editing, Visualization, Funding acquisition. \textbf{Huy Phan:} Conceptualization, Writing - Review \& Editing, Supervision. \textbf{Amir H. Ansari:} Conceptualization, Writing - Review \& Editing, Supervision. \textbf{Pascal Borz\'ee:} Resources, Writing - Review \& Editing, Data Curation. \textbf{Bertien Buyse:} Resources, Writing - Review \& Editing, Supervision. \textbf{Dries Testelmans:} Resources, Writing - Review \& Editing, Supervision. \textbf{Maarten De Vos:} Conceptualization, Methodology, Writing - Review \& Editing, Supervision, Funding acquisition.

\section*{Declaration of Competing Interest} The authors declare that they have no known competing financial interests or personal relationships that could have appeared to influence the work reported in this paper.

\section*{Acknowledgement}
This research was supported by the Research Foundation
- Flanders (FWO) [grant number 1SC2921N] and by the Flemish Government (AI Research Program).

\section*{References}

\bibliographystyle{IEEEtran.bst}
\bibliography{references}

\end{document}